\documentclass[aps, prd,floatfix, nofootinbib]{revtex4-2}
\usepackage{amsmath,amssymb,bbm}
\usepackage{tikz}
\usetikzlibrary{shapes,arrows,positioning}
\usepackage[T1]{fontenc}
\usepackage[utf8]{inputenc}
\usepackage{amsmath,amssymb,amsfonts,amstext,amsthm,amscd}
\usepackage{bm}
\usepackage{bbm}
\usepackage{bbold}
\usepackage{mathrsfs}
\usepackage{latexsym}
\usepackage{yfonts}
\usepackage{epsfig}
\usepackage{tcolorbox}
\usepackage[normalem]{ulem}
\usepackage{soul}
\usepackage{appendix}
\usepackage{lmodern}
\usepackage{xcolor}
\usepackage[
  pdfstartview=FitH,
  colorlinks=true,
  linkcolor=blue,
  anchorcolor=red,
  citecolor=magenta,
  urlcolor=blue
]{hyperref}

\setstcolor{blue}
\allowdisplaybreaks[4]
\definecolor{verde}{cmyk}{.83,.21,1,.08}

\newtheorem*{proof*}{Proof}

\newcommand{\be}{\begin{equation}}
\newcommand{\ee}{\end{equation}}
\newcommand{\beqa}{\begin{eqnarray}}
\newcommand{\eeqa}{\end{eqnarray}}
\newcommand{\eqn}[1]{(\ref{#1})}

\numberwithin{equation}{section}

\begin{document}

\title{Hamiltonian formulation of a gravity model from (A)dS Yang-Mills theory}

\author{Goffredo Chirco$^{ab}$}
 \email{goffredo.chirco@unina.it}
\author{Alfonso Lamberti$^{ab}$}
  \email{alfonso.lamberti@unina.it}
\author{Patrizia Vitale$^{abc}$}
  \email{patrizia.vitale@na.infn.it}

\affiliation{$^a$ Dipartimento di Fisica `E. Pancini', Universit\`a  di Napoli Federico II, Via Cintia 80126,  Napoli, Italy\\
$^b$ INFN, Sezione di Napoli, Italy\\
$^c$ School of Theoretical Physics Dublin Institute for Advanced Studies, 10 Burlington Road, Dublin 4, Ireland}

\begin{abstract}
We study the Hamiltonian formulation of a gravity model obtained from a Yang--Mills theory for a one-parameter family of (A)dS Lie algebras parametrized by $\alpha$, when the family of algebras is contracted to the Poincar\'e algebra in the limit $\alpha \to 0$. We derive the canonical structure and first-class constraints and analyze the resulting  algebra  in the contraction limit. In this limit, the constraints generate the residual Lorentz gauge invariance, and 
the components of the AdS potential transform as  tetrads and Lorentz connection. 
Finally, we determine the number of physical degrees of freedom, showing that in the non-propagating torsion sector---selected by a Lorentz-covariant gauge condition preserved under dynamical evolution---the theory exhibits only two propagating degrees of freedom.
\end{abstract}


\maketitle

\section{Introduction}
\label{sec1}

Dirac's theory of constrained Hamiltonian systems provides the natural framework for identifying gauge symmetries and counting physical degrees of freedom in field theory~\cite{Dirac1950}. Therefore it plays a central role both in gravity and in gauge theories of fundamental interactions.  
In this paper, we apply Dirac's analysis to the Hamiltonian formulation of the gravity model introduced in~\cite{Chirco_2025}, which arises from a suitable Yang--Mills  theory.

The model considered in~\cite{Chirco_2025} is formulated on a Minkowski background as a Yang--Mills theory whose gauge group is given by a one-parameter family of pseudo-orthogonal groups, labeled by a continuous parameter $\alpha$. Its geometric interpretation as a gravitational model emerges in a singular symmetry-breaking regime, namely in the In\"on\"u--Wigner contraction to the Poincar\'e group obtained for $\alpha\to 0$. In this limit, the gauge potential admits a decomposition in terms of a tetrad and a Lorentz connection, thus allowing for a gravity-like interpretation of the dynamics. Moreover, as shown in~\cite{Chirco_2025}, there exists a sector of the Yang--Mills dynamics in which the equations for curvature and torsion decouple, making the emergent geometric dynamics analytically tractable.

Approaches in which gravity is formulated as gauge theory 
have a long history in the literature~\cite{Kawai,Tseytlin:1981nu,Aldrovandi:1981mv,PhysRevD.21.1466,PhysRevLett.80.4851,PhysRevD.21.3269} and have more recently received renewed attention~\cite{Sobreiro:2011hb,MistrettaProkopec2023,AlexanderManton2023,Thibaut:2024uia}. Part of this renewed interest is motivated by the possibility that gauge-theoretic formulations may shed light on open conceptual and dynamical issues in gravity~\cite{weinberg1989cosmological,misner1973gravitation}, clarify its relation to symmetry-based approaches~\cite{Campiglia:2021srh}, and offer new perspectives on regimes in which General Relativity faces ultraviolet or infrared difficulties. Some of the main differences between the formulation studied here and other Yang--Mills approaches to gravity have already been discussed in~\cite{Chirco_2025}.

In this framework, the analysis of the constraint structure and of the physical degrees of freedom of the model is a necessary step further toward a geometric interpretation of the dynamics found in \cite{Chirco_2025}. This analysis is 
essential indeed, both for understanding the classical content of the model, as well as for any prospective of quantization ~\cite{ponomarev2017gauge,henneaux1992quantization}. In this sense, the first goal of the paper is to determine the primary and secondary constraints of the theory, classify them in Dirac's sense, and identify the corresponding generator of gauge transformations. To this aim, we start from the parent Yang--Mills model and compute its  first-class constraints associated with the underlying $(A)dS$ gauge symmetry.
Thereby, we focus on the behavior of the algebra of constraints in the limit  $\alpha\rightarrow 0$ and compare the result with the residual gauge invariance of the model already studied in \cite{Chirco_2025}. In particular, we clarify which components of the associated generating functional of gauge transformations survive in the limit and which conditions continue to restrict the space of admissible configurations. It is important to emphasize that the contraction limit $\alpha\rightarrow0$ does not lead to a new autonomous contracted Hamiltonian theory. Rather, the contracted dynamics is inherited from the Hamiltonian dynamics of the parent (A)dS theory. 
We  find  in particular that the constraints $\tilde{\mathcal G}_a\simeq0$ associated with the cyclicity of the translational degrees of freedom survive as descendants of the Gauss constraints of the parent theory. Although they no longer generate gauge transformations, as they are not part of the generator after the contraction, they continue to constrain the admissible configurations and therefore, as we shall see in detail, they remove four degrees of freedom in configuration space. This point is crucial for the interpretation of the reduced phase space in the limit and for the counting of the effective propagating modes.

The proposed analysis leads to two main results. First, we identify the residual Lorentz gauge symmetry governing the $\alpha\to0$ sector of the theory and determine the corresponding generating functional in phase space. Second, after imposing a Lorentz-covariant gauge condition that selects the non-propagating torsion sector, which is preserved by the Hamiltonian evolution, we find that only two effective degrees of freedom survive. This result is compatible with the proposed gravitational interpretation of the emergent dynamics in \cite{Chirco_2025} and therefore calls for further understanding.
In particular, the resulting constrained dynamics differs substantially from that of the ADM Hamiltonian formulation of General Relativity~\cite{ADM1962}. Indeed the present model is formulated in terms of first order variables and gauge-theoretic in origin: its first-class constraints are associated with internal gauge symmetry rather than directly with spacetime diffeomorphisms.\footnote{In~\cite{Chirco_2025}, diffeomorphisms can be recovered from gauge transformations when general-covariant coordinate transformations are introduced, and they emerge as symmetries of the dynamics in the vanishing-torsion sector.} It also differs from first-order Lorentz-gauge formulations of gravity, such as tetradic Palatini gravity~\cite{Palatini:1919ffw}, whose Hamiltonian structure has been analyzed for example in~\cite{Peldan1994,Rezende_2009}. In our proposal the action functional is a standard Yang-Mills one,  rather than of $BF$ type as it is the case for the above mentioned approaches; moreover,  the gauge group  is $SO(4,1)$ or $SO(3,2)$ depending on the sign of $\alpha$,  of which the Lorentz group appears only as a subgroup.

The paper is organized as follows. In Sec.~\ref{1}, we introduce the Yang--Mills Lagrangian of the model~\cite{Chirco_2025} for the one-parameter family of $(A)dS$-type Lie algebras, and we derive the canonical momenta, the primary constraints, the Hamiltonian density, and the Hamilton equations. In Sec.~\ref{1.1}, we study the preservation of the constraints and their algebra, showing that they are first class in Dirac's classification and generate gauge transformations. In Sec.~\ref{2}, we rewrite the canonical variables and the primary and secondary constraints in a form that makes their dependence on $\alpha$ explicit. 
In Sec.~\ref{3}, we analyze the algebra of constriants  in the limit $\alpha\to0$, we determine the generating functional of the residual gauge transformations, and derive the corresponding Lorentz transformation laws for the tetrads and the Lorentz connection. We then study the extended Hamiltonian in the contracted sector and perform the counting of degrees of freedom. Finally, we identify a nonpropagating torsion sector, preserved by the dynamics, where the effective number of degrees of freedom reduces to two.

\section{\label{1}Hamiltonian formulation}

The gauge  model introduced in \cite{Chirco_2025} 
is a Yang-Mills theory on Minkowski space-time $\mathcal{M} = \mathbb{R}^{(1,3)}$
with structure group $G_{\alpha}$, whose Lie algebra $\mathfrak{g}_{\alpha}$ constitutes a parametric family of (A)dS algebras depending on a dimensionless parameter $\alpha$ (see Appendix \ref{www}).
The gauge potential $\Omega$ is defined as
\begin{equation}
\Omega = \frac{1}{2} \varpi^{ab} J_{ab} + \sqrt{\frac{\lambda}{\alpha}} \vartheta^a P_a \in \Omega^1(\mathcal{M}) \otimes \mathfrak{g}_\alpha,
\end{equation}
where $\{J_{ab}, P_a\}, a=0,\dots 3$ are the Lie algebra generators while $(\varpi^{ab}, \vartheta^a)$ are the components of the connection one-form on the Lie algebra. The parameter $\lambda$ is dimensionful  with dimension of length squared, which entails that the components $\vartheta^a$ of the connection are dimensionless.
The corresponding curvature,
\begin{equation}
\mathcal{F} = d\Omega + \frac{1}{2}[\Omega, \Omega] \in \Omega^2(\mathcal{M}) \otimes \mathfrak{g}_\alpha,
\end{equation}
takes the form
\begin{equation}
\mathcal{F} = \frac{1}{2}(\mathfrak{R}^{ab} + \lambda \mathfrak{E}^{ab}) J_{ab} + \sqrt{\frac{\lambda}{\alpha}} \mathfrak{T}^a P_a,
\end{equation}
with
\begin{align}
\mathfrak{R}^{ab} &= d\varpi^{ab} + \varpi^a{}_c \wedge \varpi^{cb}, \qquad
\mathfrak{E}^{ab} = \vartheta^a \wedge \vartheta^b, \qquad
\mathfrak{T}^a = d\vartheta^a + \varpi^a{}_c \wedge \vartheta^c.
\end{align}

In natural units, the Yang--Mills action for this family of algebras has the standard form 
\begin{equation}
\label{azione}
S_{YM} = \frac{1}{2g^2} \int_{\mathcal{M}} \big(\mathcal{F} \wedge *\mathcal{F}\big)
= \int_{\mathcal{M}} d^4x\, \mathcal{L},
\end{equation}
where brackets denote the Cartan--Killing metric (see Appendix \ref{www}), and $*$ is the Hodge star defined with respect to the Minkowski metric on the base manifold.

In order to develop the Hamiltonian formulation, we follow the conventional approach for  Yang--Mills Lagrangians \cite{UtiyamaSakamoto1976}, by computing canonical momenta and constraints while   introducing  a single-index notation for the generators of the Lie algebra. In Sec.~\ref{2}, the mixed-index notation will be reintroduced  for the canonical variables, in order to make their Lie algebra components explicit.

We begin by rescaling the entire connection with the coupling constant $g$ and by adopting the single-index notation for the ten generators of the Lie algebra. The gauge connection and the curvature can then be denoted as
\begin{equation}
\Omega = \Omega^I_\mu X_I\, dx^\mu,
\qquad
\mathcal F = \frac12 F^I_{\mu\nu} X_I\, dx^\mu \wedge dx^\nu,
\end{equation}
with
\begin{equation}
F^I_{\mu\nu}
= \partial_\mu \Omega^I_\nu
- \partial_\nu \Omega^I_\mu
+ g\, f^I_{JK}\, \Omega^J_\mu \Omega^K_\nu,
\end{equation}
where $X_I$ denote the ten generators of the family $\mathfrak g_\alpha$, and $f^I_{JK}$ the corresponding structure constants in the single-index convention. The Lagrangian density in \eqref{azione} becomes
\begin{equation}
\label{lagrangiana}
\mathcal{L} = -\frac14\, F^I_{\mu\nu} F_I^{\mu\nu}.
\end{equation}

The contraction of Lie algebra indices in \eqref{lagrangiana} is performed using the Cartan--Killing metric $g_{IJ} = f^L_{IK} f^K_{JL}$. Moreover, unless explicitly stated otherwise, the passage from covariant to contravariant indices is always ensured by the Cartan--Killing metric, which is nondegenerate for simple or semisimple groups. This is the case for the family of algebras considered for $\alpha \neq 0$, for which the metric therefore also admits an inverse (see Appendix \ref{www}).

To construct the Hamiltonian formulation, we consider a foliation of the spacetime manifold $\mathcal{M}$ into spacelike hypersurfaces, such that $\mathcal{M} = \mathbb{R} \times \sigma_t$ where the hypersurfaces $\sigma_t$ are parametrized by the coordinate $x^0 = t$. The gauge fields are defined on the spatial hypersurfaces and depend on time. Accordingly, the indices $0$ and $i,j,\dots$ denote temporal and spatial coordinate indices, respectively.

The canonical momenta are computed as
\begin{align}
\Pi_I &= \frac{\partial \mathcal{L}}{\partial(\partial_0 \Omega^I_0)} = 0
\qquad \text{(primary constraints)}, \\
\Pi^i_I &= \frac{\partial \mathcal{L}}{\partial(\partial_0 \Omega^I_i)} = F_I^{i0}.
\end{align}
We observe that the canonical momenta are naturally covariant in the Lie algebra indices with respect to the Cartan--Killing metric, if we assume that the gauge fields are chosen to carry contravariant Lie algebra indices. Then, by performing the Legendre transform, we obtain the Hamiltonian density
\begin{equation}\label{ham}
\mathcal{H}
= \frac12\!\left(\Pi^I_i \Pi_I^i
+ \frac12 F^I_{ij} F_I^{ij}\right)
- \Omega_0^I\, \partial_i \Pi_I^i
- g\, f^I_{JK}\, \Omega_0^J \Omega_i^K \Pi_I^i .
\end{equation}
To write Hamilton’s equations and analyze the preservation of the primary constraints, we introduce the canonical equal-time Poisson bracket
\begin{equation}
\left\{ \Omega^I_\nu(x), \Pi^\mu_J(y) \right\}_{x_0=y_0}
= \delta^I_J\delta^\mu_\nu\, \delta^3(\bar x - \bar y).
\end{equation}
Then the Hamilton equations follow:
\begin{align}
\dot\Omega^I_i
&= \{\mathcal H, \Omega^I_i\}
= \Pi^I_i + \partial_i \Omega^I_0
- g\, f^{I}_{JK}\, \Omega^J_0 \Omega^K_i,\\
\dot\Pi_{Ii}
&= \{\mathcal H, \Pi_{Ii}\}
= \mathcal D^{(\Omega)}_j F^{\,j}_{iI}
+ f^{K}_{IJ} \Pi^i_K \Omega^J_0 .
\end{align}
where we have denoted by $\mathcal D^{(\Omega)}$ the gauge-covariant derivative with respect to the full connection. We note that, since the canonical momenta associated with $\Omega^{I}_{0}$ are primary constraints, $\Omega^{I}_{0}$ are Lagrange multipliers. In the Weyl gauge, when $\Omega^{I}_{0} = 0$, the structure of the equations generalizes to the non-Abelian case the familiar form of the Gauss constraints. 

\subsection{\label{1.1}Dirac's theory of constraints and generating functional  of gauge transformations}
In the following, we apply to the Yang-Mills Hamiltonian  \eqn{ham} the standard Dirac  analysis of constraints. 
We first determine the emergence of secondary constraints associated with the preservation of the primary ones, then we show that these are preserved on the constraint surface, so that no additional constraints arise beyond the primary and secondary ones. Finally, we analyze their algebra in order to obtain their classification.

The preservation of the primary constraints leads to secondary constraints:
\begin{equation}
\label{vincoli secondari}
\dot\Pi_{I}
=\left\{\Pi_{I},\mathcal{H}\right\}
=\partial_i\Pi^{i}_I
+g f^J_{IK}\Omega^{K}_{i}\Pi^i_{J}
=\mathcal{G}_{I}.
\end{equation}
To verify their preservation, we compute
\begin{equation}
\label{conservazione vincoli secondari}
\begin{split}
\dot{\mathcal{G}}^{I}
&=\left\{\mathcal{G}^I,\mathcal{H}\right\}
= g f^{I}_{JN}\Pi^J_i\Pi^{N\,i}
+\frac{g}{2}f^I_{JN}F^{J\,ij}F^N_{ij}
-g\Omega^J_0 g^{IL} f^K_{JL}\mathcal{G}_K .
\end{split}
\end{equation}
The first two terms of this expression vanish by construction, whereas the last term does not, but it depends linearly on the secondary constraints. Therefore  secondary constraints are preserved on the constraints surface, and so no further constraints arise beyond the primary and secondary ones. Notably, their algebra with respect to the Poisson brackets is closed:
\begin{equation}
\{\Pi_I,\Pi_{J}\}=0,\qquad
\{\Pi_I,\mathcal{G}_{J}\}=0,\qquad
\{\mathcal{G}_I,\mathcal{G}_J\}
=g f^K_{IJ}\,\mathcal{G}_K.
\end{equation}
Thus, all constraints are of first class in the Dirac classification, and therefore generate non-Abelian gauge transformations. Indeed, the  generating functional of gauge transformations can be constructed as follows \cite{Castellani:1981us}:
\begin{equation}
\label{Funzionale generatore}
K[\sigma_t]=\int_{\sigma_t} d^3\bar y\,(\beta^I\,\Pi_I+\chi^I\,\mathcal{G}_I)(y)
\end{equation}
where $\beta^I$ and $\chi^I$ are Lagrange multipliers, i.e., arbitrary functions on spacetime. With the choice
\begin{equation}
\chi^I=-\zeta^I, \quad
\beta^I=\partial_0\zeta^I+g f^I_{KJ}\Omega^K_0\zeta^J,
\end{equation}
the action of \eqref{Funzionale generatore} on the gauge connection and curvature reads
\begin{align}
\label{connection}
\delta\Omega^I_\mu
&=\{\Omega^I_\mu, K[\sigma_t]\}
=\partial_\mu\zeta^I+g f^I_{JK}\Omega^K_\mu\zeta^J,\\
\delta F^{I}_{\mu\nu}
&=\{F^{I}_{\mu\nu}, K[\sigma_t]\}
=g f^I_{JK}F^J_{\mu\nu}\zeta^K.
\end{align}
Hence the curvature is gauge covariant. Using the expression of the gauge connection in its algebra components in \eqref{connection}, we have
\begin{align}
\left\{ \vartheta^a_\mu, K[\sigma_t]\right\}&=
\left(\partial_\mu\zeta^a+g\frac{1}{2}f^a_{cd,d}\varpi^{cd}_\mu\zeta^{d}\right)\sqrt{\frac\alpha\lambda}
+g f^a_{b,cd}\vartheta^b_\mu\zeta^{cd},\\
\left\{ \frac{1}{2}\varpi^{ab}_\mu, K[\sigma_t]\right\}&=
\partial_\mu\zeta^{ab}+g\frac{1}{2}f^a_{cd,ef}\varpi^{cd}_\mu\zeta^{ef}+g\sqrt{\alpha\lambda}K^{ab}_{c,d}\vartheta^c_\mu\zeta^d.
\end{align}
As expected, we recover the gauge transformations discussed in \cite{Chirco_2025}, which, in the limit $\alpha \rightarrow 0$, reproduce the tetrad and Lorentz connection transformations. We will also discuss this aspect in the following Sec.~\ref{2} from another perspective. 

Finally, the extended Hamiltonian obtained by adding the constraints  reads
\begin{equation}
\label{H gnerale}
H_{ext}=\int_\sigma d^3\bar y\,(\mathcal{H}+\beta^I\Pi_I+\chi^I\,\mathcal{G}_I)(y),
\end{equation}
and ensures that the dynamical evolution remains on the constrained surface.

\section{\label{2}Hamiltonian dynamics for the rescaled canonical variables in the contraction limit}

In order to study the Hamiltonian in the limit $\alpha \rightarrow 0$, we rewrite all canonical variables by separating the components in the Lie algebra and making explicit their dependence on the parameters $\lambda$ and $\alpha$. This step will  allow  to take the limit $\alpha \to 0$ in a controlled manner. We also rewrite the field strengths, the secondary constraints, and the Hamiltonian by singling out the dependence on 
$\alpha$. The rescaled fields shall be indicated with  a tilde.

We have for the  components of the field strength 
\begin{equation}
F^a_{\mu\nu}
= \sqrt{\frac{\lambda}{\alpha}}\;\tilde F^a_{\mu\nu}
= \sqrt{\frac{\lambda}{\alpha}}\;\mathfrak T^a_{\mu\nu},\qquad
F^{ab}_{\mu\nu}
= \frac12\,\tilde F^{ab}_{\mu\nu}
= \frac12(\mathfrak R^{ab}_{\mu\nu}+\lambda\mathfrak E^{ab}_{\mu\nu}),
\end{equation}
while the canonical momenta are given by
\begin{align}
\label{P1}
\Pi_a
&= \frac{\partial\mathcal L}{\partial(\partial_0\Omega^a_0)}
= \sqrt{\frac{\alpha}{\lambda}}
\frac{\partial\mathcal L}{\partial(\partial_0\vartheta^a_0)}
= \sqrt{\frac{\alpha}{\lambda}}\,\tilde\pi_a ,\\
\label{P2}
\Pi_{ab}
&= \frac{\partial\mathcal L}{\partial(\partial_0\Omega^{ab}_0)}
= 2\frac{\partial\mathcal L}{\partial(\partial_0\varpi^{ab}_0)}
= 2\tilde\pi_{ab},\\
\label{P3}
\Pi^i_a
&= \sqrt{\frac{\alpha}{\lambda}}\,\tilde\pi^i_a,\\
\label{P4}
\Pi^i_{ab}
&= 2\tilde\pi^i_{ab}.
\end{align}
Substituting the previous expressions into the secondary constraint equations \eqref{vincoli secondari}, we obtain
\begin{align}
\label{G1}
\mathcal G_a
&= \sqrt{\frac{\alpha}{\lambda}}
\left(
\partial_i\tilde\pi^i_a
+ g\lambda K^{cd}_{a,b}\vartheta^b_i\tilde\pi^i_{cd}
+ \frac{g}{2}f^d_{a,bc}\varpi^{bc}_i\tilde\pi^i_d
\right)
= \sqrt{\frac{\alpha}{\lambda}}\,\tilde{\mathcal G}_a ,\\
\label{G2}
\mathcal G_{ab}
&= 2\left(
\partial_i\tilde\pi^i_{ab}
+ \frac12 g f^{ef}_{ab,cd}\varpi^{cd}_i\tilde\pi^i_{ef}
+ \frac12 g f^{e}_{ab,d}\vartheta^{d}_i\tilde\pi^i_e
\right)
= 2\tilde{\mathcal G}_{ab}.
\end{align}
We therefore find that the canonical momenta and the secondary constraints associated with the $P^a$ sector of the Lie algebra depend on $\alpha$ in a natural way. This fact will play a crucial role when we study the limit in which $\alpha$ tends to zero.

Let us write the extended Hamiltonian density by decomposing it into Lie algebra components:
\begin{equation}
\begin{aligned}
\label{H1}
\mathcal H_{ext}
&= \frac12\left(
\Pi^a_i\Pi_a^i
+ \Pi^{ab}_i\Pi_{ab}^i
+ \frac12 F^a_{ij}F_a^{ij}
+ \frac12 F^{ab}_{ij} F_{ab}^{ij}
\right)
\\
&\quad
- \Omega^a_0\mathcal G_a
- \Omega^{ab}_0\mathcal G_{ab}+ \beta^a \Pi_a
+ \beta^{ab}\Pi_{ab}
+ \chi^a\mathcal G_a
+ \chi^{ab}\mathcal G_{ab},
\end{aligned}
\end{equation}
Recalling that, with respect to Lie algebra indices, the canonical momenta carry a covariant index, whereas the field strength tensor and the gauge fields carry a contravariant one\footnote{In \eqref{H1}, as in \eqref{H gnerale}, the Lagrange multipliers are chosen with contravariant Lie-algebra indices.}
we make the contractions in \eqref{H1} explicit by means of the Cartan--Killing metric in its covariant and contravariant forms. 
Hence one obtains
\begin{equation}
\label{Htot}
\begin{aligned}
\mathcal{H}_{ext}
&=\frac{1}{2}(G^{a,b}\Pi_{ib}\Pi_a^i+G^{ab,cd}\Pi_{cdi}\Pi_{ab}^i+\frac{1}{2}G_{a,b}{F}^a_{ij} F^{b\,ij}+\frac{1}{2}G_{ab,cd}F^{ab}_{ij} F^{cd\,ij})
\\
&\quad
-\Omega^{a}_0\mathcal{G}_a-\Omega^{ab}_0\mathcal{G}_{ab}
+\beta^a\Pi_a+\beta^{ab}\Pi_{ab}+\chi^a\mathcal{G}_a+\chi^{ab}\mathcal G_{ab}.
\end{aligned}
\end{equation}
with $G$ the Cartan-Killing metric detailed in App.~\ref{www}. 
Substituting \eqref{InvMetric1} and \eqref{InvMetric2} and using the previous rescaled canonical variables and constraints, Eq.  \eqref{Htot} becomes
\begin{equation}
\label{H tot finale}
\mathcal{H}_{ext}
=\mathcal{H}
+ \sqrt{\frac{\alpha}{\lambda}}\,\beta^a\tilde\pi_a
+ 2\beta^{ab}\tilde\pi_{ab}
+ \sqrt{\frac{\alpha}{\lambda}}\,\chi^a\,\tilde{\mathcal G}_a
+ 2\chi^{ab}\tilde{\mathcal G}_{ab}.
\end{equation}
where the Hamiltonian $\mathcal{H}$ defined in \eqref{ham} now reads
\begin{align}
\mathcal H
&= \tilde\pi^{ab}_i\tilde\pi_{ab}^i
+ \frac{1}{2\lambda}\tilde\pi^{a}_i\tilde\pi_{a}^i
+ \frac{\lambda}{4}\tilde F^a_{ij}\tilde F_a^{ij}
+ \frac12 \tilde F^{ab}_{ij}\tilde F_{ab}^{ij} - \vartheta^{a}_0\tilde{\mathcal{G}}_a
- \varpi^{ab}_0\tilde{\mathcal{G}}_{ab},\label{hamnaive}\\[2mm] 
\tilde{\mathcal G}_a
&= \partial_i\tilde\pi^i_a
+ g\lambda K^{cd}_{a,b}\vartheta^b_i\tilde\pi^i_{cd}
+ \frac{g}{2}f^d_{a,bc}\varpi^{bc}_i\tilde\pi^i_d,\\[2mm]
\tilde{\mathcal G}_{ab}
&= \partial_i\tilde\pi^i_{ab}
+ \frac12 g f^{ef}_{ab,cd}\varpi^{cd}_i\tilde\pi^i_{ef}
+ \frac12 g f^{e}_{ab,d}\vartheta^{d}_i\tilde\pi^i_e.
\end{align}
We find that the na\"ive Hamiltonian is independent of $\alpha$, as expected, since the Lagrangian is independent of $\alpha$ by construction, as explained in \cite{Chirco_2025}. The Hamiltonian dynamics for the  rescaled canonical variables reads then
\begin{align}
\dot\vartheta^a_i
&= \frac{1}{\lambda}\,\tilde\pi^a_i
+ \mathcal D^\varpi \vartheta^a_0
- g\,f^{a}{}_{cb,d}\,\vartheta^d_i\,\varpi^{cb}_0,\\
\dot\varpi^{ab}_i
&= 4\,\tilde\pi^{ab}_i
+ \mathcal D_i^\varpi \varpi^{ab}_0
- 2g\lambda\,K^{ab}{}_{c,d}\,\vartheta^d_i\,\vartheta^c_0,\\
\dot{\tilde{\pi}}^i_a
&= \frac{1}{\lambda}\mathcal D^\varpi_j\,\mathfrak{T}^{ji}_{a}
+ \frac{2g}{\lambda}\left(\mathfrak{R}^{ij}_{ac}\vartheta^c_j + \lambda\,\mathfrak{E}^{ij}_{ac}\vartheta^c_j\right)
+ \frac{2g}{\lambda}\,\tilde{\pi}^i_{ac}\vartheta^c_0
+ \frac{1}{2\lambda}f^d{}_{a,cd}\,\tilde{\pi}^i_d\,\varpi^{cd}_0,\\
\dot{\tilde{\pi}}^i_{ab}
&= \mathcal{D}^\varpi_j(\mathfrak{R}^{ij}_{ab} + \lambda\mathfrak{E}^{ij}_{ab})
+ \frac{g\lambda}{2}f^d{}_{ab,c}\,\vartheta^c_j\,\mathfrak{T}^{ij}_d
+ \frac{g}{2}f^d{}_{ab,c}\tilde{\pi}^i_d\vartheta^c_0
+ \frac{g}{2}f^{ef}{}_{ab,cd}\,\tilde{\pi}^{i}_{ef}\varpi^{cd}_0.
\end{align}
Finally, in the Weyl gauge $\vartheta^a_0=0$, $\varpi^{ab}_0=0$, the Hamilton equations for the spatial momenta manifestly reproduce the structure of the equations of motion in \cite{Chirco_2025}, involving only the canonical variables on spacelike hypersurfaces.

\section{\label{3} The algebra of constraints in the limit \texorpdfstring{$\alpha \to 0$}{alpha -> 0}}
After isolating the parameter $\alpha$, we take the limit $\alpha \to 0$ for primary and secondary constraints. We obtain
\begin{equation}
\begin{aligned}
\lim_{\alpha \to 0}\Pi_a &= \lim_{\alpha \to 0}\sqrt{\frac{\alpha}{\lambda}}\,\tilde\pi_a = 0 ,
\qquad
\lim_{\alpha \to 0}\Pi_{ab} = \lim_{\alpha \to 0} 2\tilde\pi_{ab} = 2\tilde\pi_{ab} ,
\\[4pt]
\lim_{\alpha\to 0}\mathcal G_a &= \lim_{\alpha\to 0}\sqrt{\frac{\alpha}{\lambda}}\,\tilde{\mathcal{G}}_a = 0 ,
\qquad
\lim_{\alpha\to 0}\mathcal G_{ab} = 2\tilde{\mathcal{G}}_{ab} .
\end{aligned}
\end{equation}
We observe that only the  constraints associated with Lorentz gauge invariance survive the limit, and the constraint algebra becomes
\begin{equation}
\label{L algebra}
\{\tilde\pi_{ab},\tilde\pi_{cd}\}=0 ,
\qquad
\{\tilde\pi_{ab},\tilde{\mathcal{G}}_{cd}\}=0,
\qquad
\{\tilde{\mathcal G}_{ab},\tilde{\mathcal G}_{cd}\}
= g\,f^{ef}_{ab,cd}\,\tilde{\mathcal {G}}_{ef} .
\end{equation}
As for the  the generating functional of gauge transformations, we get,  in the limit $\alpha\to 0$
\begin{equation}
\label{K0}
\lim_{\alpha \to 0} K[\sigma_t]
= \int_\sigma d^3\bar y \left(
2\beta^{ab}\tilde\pi_{ab}
+ 2\chi^{ab}\tilde{\mathcal{G}}_{ab}\right).
\end{equation}
That is,
\begin{equation}
\lim_{\alpha \to 0} K[\sigma_t]
= \int_\sigma d^3\bar y \left[
2\beta^{ab}\tilde\pi_{ab}
+ 2\chi^{ab}\Big(
\partial_i\tilde\pi^i_{ab}
+ \frac{1}{2}g f^{ef}_{ab,cd}\varpi^{cd}_i\tilde \pi^i_{ef}
+ \frac{1}{2}g f^e_{ab,d}\vartheta^d_i\tilde \pi^i_e
\Big)\right]
\end{equation}
with multipliers
\begin{equation}
2\chi^{ab}=-\zeta^{ab},\qquad
2\beta^{ab}=\partial_0\zeta^{ab}+g f^{ab}_{cd,ef}\tfrac12\varpi^{cd}_0\zeta^{ef}.
\end{equation}
We can observe that the generating functional of gauge transformations in the limit generates the correct Lorentz transformations for the fields:
\begin{equation}
\begin{aligned}
\delta\vartheta^a_i
&= \left\{\vartheta^a_i,\lim_{\alpha \to 0} K[\sigma_t]\right\}
= \frac{1}{2}g f^a_{d,ef}\,\zeta^{ef}\,\vartheta^d_i ,
\\
\delta\varpi^{ab}_i
&= \left\{\varpi^{ab}_i,\lim_{\alpha \to 0} K[\sigma_t]\right\}
= \partial_i\zeta^{ab}
+ \frac12 g f^{ab}_{ef,cd}\,\varpi^{cd}_i\,\zeta^{ef} .
\end{aligned}
\end{equation}
These transformations are consistent with those of the tetrads  $\vartheta^a$, that transform as  vectors, and the Lorentz connection $\varpi^{ab}$ that transforms inhomogeneously. Notably, the extended Hamiltonian in the limit $\alpha \rightarrow 0$ takes the form
\begin{equation}
\label{H limi}
\lim_{\alpha\to 0}\mathcal{H}_{ext}
=\mathcal{H}
+ 2\beta^{ab}\tilde \pi_{ab}
+ 2\chi^{ab}\tilde{\mathcal G}_{ab}.
\end{equation}
Finally let us come to  the counting of degrees of freedom. As for  the  starting dimension of   phase space we have:
\begin{equation}
\Omega^I_{\mu}\,\,:\,\,40 \qquad \longrightarrow \qquad (\Omega^I_{\mu}, \Pi^I_{\mu})\,\,:\,\,80 .
\end{equation}
Before taking the  limit $\alpha\rightarrow0$,  we have 20 first-class constraints of which 12 related to Lorentz invariance and 8  to the non-commuting translations:
\begin{equation}
\tilde{\pi}_a,\quad \tilde{\pi}_{ab},\quad \tilde{\mathcal{G}}_a,\quad \tilde{\mathcal{G}}_{ab}.
\end{equation}
They generate gauge transformations, thereby reducing the number of degrees of freedom in the configuration space to
\begin{equation}
20 = \frac{80 - 12 \times 2 - 8 \times 2}{2}.
\end{equation}
Upon taking the limit  $\alpha\rightarrow0$, the set of constraints that survive in the extended Hamiltonian  \eqref{H limi} is given by
\begin{equation}
\tilde{\pi}_{ab},\quad 
\tilde{\mathcal{G}}_{ab}.
\end{equation}
Indeed, from the analysis of the evolution of $\tilde{\pi}_a$ and $\vartheta^a_0$, these variables become constants of motion. The constraints $\tilde{\pi}_{ab}$ and $\tilde{\mathcal{G}}_{ab}$ remain first class and continue to generate Lorentz gauge transformations; in contrast, the constraints $\tilde{\mathcal{G}}_a$ still contributing to the naive Hamiltonian in \eqref{hamnaive}, although remaining conserved,\footnote{This can be proved since the Poisson bracket between \eqref{H limi} and these constraints still yields a combination of constraints, as shown in \eqref{conservazione vincoli secondari}.} no longer contribute to the generation of gauge symmetries, 
as they do not appear in the generating functional \eqref{K0}. As a consequence, the reduction of the initial gauge symmetry to Lorentz gauge symmetry, in the limit $\alpha\rightarrow0$, allows the release of degrees of freedom, leading to a total number of degrees of freedom equal to
\begin{equation}
26 = \frac{80 - 12 \times 2 - 4}{2}
\end{equation}
in configuration space.

This summarizes the counting of degrees of freedom in configuration space after taking $\alpha\to0$. The limit  effectively breaks the original gauge symmetry to the one generated by the sole Lorentz subalgebra, thus establishing a relation with the Local Lorentz invariance of first order gravity models.   However, as already pointed out  in \cite{Chirco_2025}, the dynamics sofar obtained is considerably more general than that of  first-order tetradic approaches, due to the presence of dynamical torsion terms. In our model, by only specifying the sources  is not sufficient to guarantee a decoupling of the equations and to recover an algebraic relation between torsion and spin. Nevertheless, as shown in \cite{Chirco_2025} and summarized in Appendix \ref{ttt}, there exists a sector of the gauge dynamics, defined by a Lorentz-covariant gauge condition, where the dynamics decouples and an algebraic relation between torsion and spin is restored. 

It is precisely within this sector that a  connection with  first-order gravity models can be established, and the equations become analytically tractable. Let us restrict  therefore to the case of nonpropagating torsion as  (see Appendix \ref{ttt}). This choice   imposes 24 additional constraints ($4\times 6$ are the components of the torsion field) to the counting of the  degrees of freedom. We have then
\begin{equation}
\mathcal D^\mu_{\varpi}\mathcal{F}_{[\mu\nu]}^{ab}=0
\;\rightarrow\;
{T}^{a}{}_{[\mu\nu]}=\chi S^a_{\mu\nu},
\qquad 26 - 24 = 2.
\end{equation}
We obtain two effective degrees of freedom in configuration space. This sector is preserved by the dynamics on the solution space, as can be verified by taking the Poisson bracket with  the extended Hamiltonian
\begin{equation}
\{ \mathcal H_{ext}, \mathcal D_{\varpi}^{\mu}  \mathcal F^{ab}_{[\mu\nu]} \}_{\text{on-shell}} =0,
\end{equation}
 and so no additional constraints are generated.

\section{Discussion}
We have studied the Hamiltonian formulation of the gravity model proposed in \cite{Chirco_2025}, as the limit $\alpha \to0$ of a one-parameter family of Yang-Mills models with (A)dS gauge symmetry. We have first  analyzed the canonical structure and derived the first-class constraints of the  Yang--Mills theory, and then  performed the  limit $\alpha \to 0$ for the algebra of constraints and the corresponding gauge transformations. In this limit, the analysis reveals that the constraints are associated with residual Lorentz gauge invariance. Additionally, the generator of gauge transformations yields field transformations that are consistent with those of tetrads and the Lorentz connection. 
{Furthermore, after imposing a Lorentz-covariant gauge condition that selects the non-propagating torsion sector,  we find that only two effective degrees of freedom survive. Thus, the proposed analysis of the constraints structure  supports the gravitational interpretation of the emergent dynamics in (A)dS Yang–Mills and highlights the need for further understanding. }

At the gauge level, before taking the limit  $\alpha \rightarrow 0$, it is important to note that the model corresponds to a standard Yang–Mills theory for a simple but noncompact group. At the quantum level, the noncompactness of the structure group in  gauge theories is known to have important implications, particularly the potential emergence of unitarity violations—namely, negative-norm states and vacuum instability.  These issues stem from the indefiniteness of the Killing metric (see for example ~\cite{AlexanderManton2023, MargolinStrazhev1992} and refs. therein where the problem has been addressed). {From the perspective of using the proposed model to study the quantization of geometric dynamics,} the lack of adequate mechanisms could compromise the consistency of the theory.

In the context of the model \cite{Chirco_2025}, the identification of such pathologies is not immediate from the Hamiltonian perspective. Our analysis has been carried out at the classical level. However, at the quantum level, a gauge-fixing term becomes unavoidable and must be incorporated into the formulation. Consequently, a consistent treatment requires extending the analysis to this setting. Moreover, even upon imposing additional conditions (such as those leading to non-propagating torsion) and adopting a second-order formulation in the torsionless case, the resulting reduction in the number of degrees of freedom does not automatically guarantee the physical consistency of the emergent modes. In particular, the effective theory governing the dynamics of the metric field must be investigated.

For these reasons, the canonical analysis should be supplemented by complementary approaches, including a perturbative study of the propagator, a BRST cohomological analysis of the physical state space, and a thorough examination of the kinetic operator to ensure the absence of instabilities. Similar considerations apply to the effective theory for the emergent metric field in the regime where the formulation reduces to second order. 

These issues, specific of the quantization of the theory, as well as a deeper understanding of the dynamics described by the model, with and without the assumption of nonpropagating torsion, deserve further investigation and shall be addressed in future work. 

\section*{Acknowledgments}
The authors acknowledge support from the INFN Iniziativa Specifica GeoSymQFT and from  the European COST Action CaLISTA CA21109.  P.V. acknowledges support  from  the PNRR MUR Project No. CN 00000013-ICSC.

\bibliography{references}

\appendix

\section{The (A)dS groups}
\label{www}

In this Appendix we summarize some notations that were discussed in more detail in \cite{Chirco_2025}. The family $\mathfrak{g}_\alpha$ of (A)dS algebras depending on a continuous parameter $\alpha$, discussed in Sec.~\ref{1}, has generators satisfying the following Lie brackets:
\begin{equation}
\label{P-algebra}
\begin{split}
[P_a,P_b]&:=\alpha\, K_{a,b}^{ef}J_{ef},
\\
[P_a,J_{bc}]&=\boldsymbol{\eta}_{ab}P_c-\boldsymbol{\eta}_{ac}P_b:=f^e_{a,bc}P_e,
\\
[J_{ab},J_{cd}]&=\boldsymbol{\eta}_{ad} J_{bc}
- \boldsymbol{\eta}_{ac} J_{bd}
+ \boldsymbol{\eta}_{bc} J_{ad}
- \boldsymbol{\eta}_{bd} J_{ac}
:=f^{ef}_{ab,cd}J_{ef}.
\end{split}
\end{equation}
The generators are $J_{ab}$ and $P_a$, with $a,b=0,\dots,3$, and $K^{cd}_{a,b}=\frac{1}{2}\left(\delta^c_a\delta^d_b-\delta^d_a\delta^c_b\right)$, while $\boldsymbol{\eta}_{ab}$ is the four-dimensional Minkowski metric with signature $(+,-,-,-)$. The parameter $\alpha$ can take positive, negative, or zero values, allowing the identification of the algebras $\mathfrak{so}(1,4)$, $\mathfrak{so}(2,3)$, and $\mathfrak{iso}(1,3)$, respectively. The limit $\alpha \to 0$ corresponds to the In\"on\"u--Wigner contraction~\cite{Inonu}, which produces the Poincar\'e algebra $\mathfrak{iso}(1,3)$ from either $\mathfrak{so}(1,4)$ or $\mathfrak{so}(2,3)$.

Within our parametrization the Cartan--Killing metric can be checked to be
\begin{align}
G_{a,b} &= \alpha\,\boldsymbol{\eta}_{ab}, \label{FinaleJJ} \\
G_{ab,cd} &= \boldsymbol{\eta}_{ac}\boldsymbol{\eta}_{bd} - \boldsymbol{\eta}_{bc}\boldsymbol{\eta}_{ad}. \label{FinaleJP}
\end{align}
It becomes degenerate in the limit $\alpha \to 0$, i.e., when the algebra reduces to $\mathfrak{iso}(1,3)$. This metric is diagonal in the algebra and the inverse metric is given by
\begin{align}
G^{a,b} &= \alpha^{-1}\boldsymbol{\eta}^{ab}, \label{InvMetric1} \\
G^{ab,cd} &= \frac{1}{4} (\boldsymbol{\eta}^{ac}\boldsymbol{\eta}^{bd} - \boldsymbol{\eta}^{bc}\boldsymbol{\eta}^{ad}). \label{InvMetric2}
\end{align}
This allows one to define a relation between covariant and contravariant objects in the algebra.

\section{Gauge dynamics}
\label{ttt}
We shortly review here the gauge dynamics derived from the action \eqref{azione}. We refer to \cite{Chirco_2025} for details. On considering the variation of the action with respect to the gauge   connection and by adding to the action a gauge invariant source term one obtains 
\[
\frac{\hbar}{2 g^2}\mathcal{D}_\Omega *\mathcal{F} = -J ,
\]
with $J$ the current three-form given by
\be J= \frac{1}{2} \Sigma^{ab}J_{ab} + \sqrt{\frac{1}{\lambda\alpha}}\Theta^a P_a
\ee
and $\Sigma^{ab}, \Theta^a$ respectively the spin and energy-momentum components (see \cite{Chirco_2025} for details). By decomposing in  the Lie algebra we have 
\begin{equation*}
\begin{split}
\mathcal{D}_\varpi(*\mathfrak{T}^a) + \vartheta_b \wedge *\mathcal{F}^{ab} + 
&= \chi\, \Theta^a, \\
\mathcal{D}_\varpi(*\mathcal{F}^{ab})+ 2\, \vartheta^a \wedge *\mathfrak{T}^b &= \chi\, \Sigma^{ab}.
\end{split}
\end{equation*}
with $\chi= -2 g^2/\hbar\lambda$. By considering the Lorentz-invariant sector $\mathcal{D}_\varpi(*\mathcal{F}^{ab})=0$ the would be torsion is directly connected to the spin source and therefore nonpropagating. 
Performing the Hodge dual with respect to the Minkowski metric, and taking the limit $\alpha \to 0$, where the fields can be consistently interpreted as   tetrad  and Lorentz connection (see Sec.~4 of \cite{Chirco_2025}), the previous equations become
\begin{equation*}
\begin{split}
\mathcal{D}_\mu^\omega\, T_{[\beta\nu]}^a\, \eta^{\mu\nu} + R^{ab}_{[\beta\nu]}\, \theta_{b\mu}\, \eta^{\mu\nu}
+ \lambda\left(\theta_{b\mu} \theta^a_\beta \theta^b_\nu
- \theta_{b\mu} \theta^a_\nu \theta^b_\beta \right) \eta^{\mu\nu} &= \chi\, \mathscr{T}^a_\beta, \\
2\, \theta^a_\mu\, T^b_{[\beta\nu]}\, \eta^{\mu\nu} &= \chi\, \mathcal{S}_\beta^{ab}.
\end{split}
\end{equation*}
where
\begin{align*}
R^{ab}_{[\mu\nu]} &= \partial_{\mu} \omega^{ab}_{\nu} - \partial_{\nu} \omega^{ab}_{\mu}
+ \omega^{a}_{\mu c}\, \omega^{cb}_{\nu} - \omega^{a}_{\nu c}\, \omega^{cb}_{\mu}, \\
T^a_{[\mu\nu]} &= \partial_{\mu} \theta^a_{\nu} - \partial_{\nu} \theta^a_{\mu}
+ \omega^a_{\mu c} \theta^c_{\nu} - \omega^a_{\nu c} \theta^c_{\mu}
\end{align*}
while $\mathscr{T}^a_\beta$ and $\mathcal{S}^{ab}_\beta$ are, respectively, the components of the energy--momentum and  spin three forms $\Theta^a$ and $\Sigma^{ab}$. 



\end{document}